\documentclass[twocolumn,preprintnumbers,amsmath,amssymb,aps,physrev,superscriptaddress]{revtex4-2} 

\usepackage{amsmath} 
\usepackage[english]{babel} 
\usepackage{graphicx} 
\usepackage{listings} 
\usepackage{soul}
\usepackage{ulem}
\usepackage{color}
\usepackage{longtable}

\begin{document}

\title{Molecular origin of the two-step mechanism of gellan aggregation}

\author{Letizia Tavagnacco}
\email[Corresponding author: ]{letizia.tavagnacco@cnr.it}
\affiliation{CNR-ISC and Department of Physics, Sapienza University of Rome, Piazzale A. Moro 2, 00185, Rome, Italy.}
\author{Ester Chiessi}
\affiliation{Department of Chemical Sciences and Technologies, University of Rome Tor Vergata, Via della Ricerca Scientifica I, 00133 Rome, Italy.}
\author{Leonardo Severini}
\affiliation{Department of Chemical Sciences and Technologies, University of Rome Tor Vergata, Via della Ricerca Scientifica I, 00133 Rome, Italy.}
\author{Silvia Franco}
\affiliation{CNR-ISC and Department of Physics, Sapienza University of Rome, Piazzale A. Moro 2, 00185, Rome, Italy.}
\author{Elena Buratti}
\affiliation{CNR-ISC and Department of Physics, Sapienza University of Rome, Piazzale A. Moro 2, 00185, Rome, Italy.}
\author{Angela Capocefalo}
\affiliation{CNR-ISC and Department of Physics, Sapienza University of Rome, Piazzale A. Moro 2, 00185, Rome, Italy.}
\author{Francesco Brasili}
\affiliation{CNR-ISC and Department of Physics, Sapienza University of Rome, Piazzale A. Moro 2, 00185, Rome, Italy.}
\author{Adriano Mosca Conte}
\affiliation{CNR-ISC and Department of Physics, Sapienza University of Rome, Piazzale A. Moro 2, 00185, Rome, Italy.}
\author{Mauro Missori}
\affiliation{CNR-ISC and Department of Physics, Sapienza University of Rome, Piazzale A. Moro 2, 00185, Rome, Italy.}
\author{Roberta Angelini}
\affiliation{CNR-ISC and Department of Physics, Sapienza University of Rome, Piazzale A. Moro 2, 00185, Rome, Italy.}
\author{Simona Sennato}
\affiliation{CNR-ISC and Department of Physics, Sapienza University of Rome, Piazzale A. Moro 2, 00185, Rome, Italy.}
\author{Claudia Mazzuca}
\affiliation{Department of Chemical Sciences and Technologies, University of Rome Tor Vergata, Via della Ricerca Scientifica I, 00133 Rome, Italy.}
\author{Emanuela Zaccarelli}
\email[Corresponding author: ]{emanuela.zaccarelli@cnr.it}
\affiliation{CNR-ISC and Department of Physics, Sapienza University of Rome, Piazzale A. Moro 2, 00185, Rome, Italy.}


\begin{abstract}
Among hydrocolloids, gellan is one of the most used anionic polysaccharides, because of its capability of forming mechanically stable gels at relatively low concentrations. Despite its long-standing use and importance, the gellan aggregation mechanism is still not presently understood at the microscopic level due to the lack of atomistic information. Here we will fill this gap by reporting molecular dynamics simulations of gellan chains at different polymer and salt contents, being able to unveil the occurrence of the two steps in the process, in agreement with existing hypotheses. At first, the formation of double helices takes place, followed by the aggregation into super-structures. For both steps, the role of bivalent cations appears to be crucial, as also shown by rheology and atomic force microscopy measurements: they not only facilitate the junction of the chains into double helices, but also promote through bridging their arrangement into larger aggregates. On the other hand, monovalent cations have a much more reduced role, making it possible to form double helices only at very high salt content and not actively participating in the formation of gels. Our simulations thus offer the first complete microscopic overview of gellan aggregation and will be important for future use of gellan-based systems for a wide variety of applications, ranging from food science to art restoration.

\end{abstract}

\maketitle


\section{Introduction}
Gellan is an anionic microbial exopolysaccharide that has gained increasing interest in the pharmaceutical, cosmetics, and food industry because of its functional and mechanical properties~\cite{osmalek2014application,matsusaki2019fabrication,muller2020development,liu2020highly,villarreal2021fabrication}. It is a water-soluble, bio-compatible, atoxic, biodegradable and also chemically stable at high temperature polymer~\cite{chandrasekaran1995molecular,iurciuc2016gellan}. Gellan is characterized by a linear structure consisting of tetrasaccharide repeating units, i.e. D-Glc($\beta1\rightarrow 4$)D-GlcA($\beta1\rightarrow 4$)D-Glc($\beta1\rightarrow 4$)L-Rha($\alpha1\rightarrow 3$)~\cite{jansson1983structural,o1983structure}. As a consequence, each repeating unit contains one carboxyl group belonging to the glucuronic acid that, when ionized in aqueous solution, confers the anionic character to the polysaccharide. The native polysaccharide also includes a L-glyceril and an acetyl group linked to the glucose moiety, that are industrially removed to produce a deacetylated polymer, commonly known as gellan gum. The structure of polycrystalline fibers of the lithium salt of the deacetylated gellan was determined by X-Ray diffraction as consisting of a double helix formed by two left-handed, threefold helical chains~\cite{chandrasekaran1988crystal}, and lately confirmed also for the potassium form~\cite{chandrasekaran1988cation}.

The most important property of gellan, which is at the origin of its widespread use, is its capability of forming thermo-reversible transparent gels by cooling aqueous solutions containing cations, even at low polymer concentrations. The gellan sol-gel transition is an exothermic process~\cite{miyoshi1994gel} that can occur in a range of temperature between 30 and 50$^{\circ}$C, depending on the specific nature of the cation, polymer concentrations and presence of cosolutes~\cite{moritaka1992effects,watase1993effect,miyoshi1996rheological,perez2012gelation}. In particular, it was observed that divalent cations determine a higher gelling temperature with respect to the monovalent counterpart~\cite{tang1997gelling}. Cations were also found to affect the microstructure and the water holding capacity of gellan gels~\cite{mao2001water}, as well as gel strength~\cite{grasdalen1987gelation}. In addition, Atomic Force Microscopy (AFM) measurements revealed that the network structures of gellan gels obtained with sodium chloride is more heterogeneous than those formed in potassium chloride~\cite{funami2009molecular}.

Another appealing property of gellan is its suitability as innovative agent for wet cleaning treatments in the restoration of paper artworks~\cite{mazzuca2014gellan}. In a recent work~\cite{di2020gellan}, it was also demonstrated that microgels based on gellan, i.e. micro-scale particles internally made by a gellan cross-linked network, offer several advantages for paper cleaning with respect to more established procedure based on wet cleaning or hydrogels systems. Indeed, paper is a complex material mainly formed by interwoven cellulose fibers, whose composition and structure can vary depending on the production process, that deteriorates with time due to environmental conditions~\cite{conte2012role,corsaro2013molecular,missori2019non}. Owing to the reduced size, microgels based on gellan can better penetrate the porous structure of paper and remove pollution and degradation materials in a shorter time~\cite{di2020gellan}. In addition, the softness of microgels suspensions can be adapted to the irregular surface of paper artworks, thus providing an higher efficiency. It is interesting to note that gellan hydrogels employed for paper cleaning are formed in the presence of calcium acetate~\cite{mazzuca2014gellan}, while gellan microgels are prepared by applying external shear upon the addition of sodium chloride~\cite{caggioni2007rheology}. It is thus important, also in the context of its cultural heritage applications, to clarify the role of cations on gellan aggregation.

The mechanism of gelation of gellan has been the source of many investigations. The main hypothesis is that of a two-step process~\cite{morris2012gelation,iurciuc2016gellan}: in the first step, the gellan chains form double helix structures, even in the absence of gel promoting cations~\cite{grinberg2003thermodynamics}; then, in a second step,  the double helixes form cation-mediated aggregates that compose the gel network. However, a recent work based on statistical analysis of AFM images has also proposed that the first ordered state is a single helix~\cite{diener2020rigid}. Moreover, different interpretations of the two-step process still exist~\cite{morris2012gelation}. Based on light scattering experiments, Gunning et al.~\cite{gunning1990light} proposed a model in which the first step consists in the formation of fibrils, produced by double helix formation between ends of neighbouring molecules, later followed by lateral crystallization of the fibrils mediated by cations, which ultimately leads to the formation of gels. On the other hand, by using differential scanning calorimetry and rheology, Robinson et al.~\cite{robinson1991conformation} proposed that gelation occurs first trough a conversion from disordered coils into double helix structures, then only some of these double helixes form cation-mediated aggregates that compose the gel network. Hence, the second scenario includes a larger degree of disorder in the network, for the presence of disordered chains or unaggregated double helices, compatible with rheological signatures of weak gels.

Importantly, such a dispute has not been settled in the last thirty years, leaving unanswered (i) if and how the double helix forms and (ii) the role of cations in gel formation. To this aim, molecular dynamics simulations are a valuable tool since they allow for the investigation of these processes at the molecular scale. In order to perform this study, a force field for the gellan chains is needed, because not available in the literature, except for a recent study which however focused on acetylated gellan~\cite{jamil2019dynamics}, that is not the one most widely used in applications because of its limited gelation ability~\cite{iurciuc2016gellan}.

Motivated by these reasons, in this work, we propose a new force field for (deacylated) gellan and perform extensive atomistic molecular dynamics simulations of chains under a variety of conditions, including the addition of monovalent and bivalent salts. Comparing numerical results with atomic force microscopy and rheological measurements, we fully characterize the microscopic process of gellan aggregation and unveil the role played by the cations. Our findings provide a microscopic confirmation that the aggregation of gellan occurs in two steps. First, a conformational transition takes place from single polysaccharide chains to double helix structures, which can happen with and without the mediation of the cations. Next, simulations reveal a different behavior of gelation in the presence of  monovalent and bivalent cations. Indeed, only the latter are able to promote the association of double helix structures into super-aggregates through the complexation of carboxylate groups belonging to different double helix structures. Instead, in the presence of monovalent salt at the same charge content, we do not observe the second step within the time window of our simulations. These results provide evidence of the underlying mechanisms occurring in two-step gellan aggregation and unveil the specific role of bivalent cations, which clearly enhance the formation of a gel.

\section{Sample preparation}
Deacetylated gellan(KELCOGEL CG-LA), also known as gellan gum and hereafter referred to as gellan, was purchased from CP Kelco (Atlanta Georgia, USA). Calcium acetate and sodium chloride are from Merck (Merck KGaA, Darmstadt, Germany). Ultrapure water (MilliQ, Millipore, Billerica, MA, USA) was used in the preparation of solutions (resistivity 18.2 M$\Omega$ cm at T=25$^{\circ}$C). Three different systems were compared: gellan hydrogels without added salts ($G_{pure}$), hydrogels with sodium chloride ($G_{Na}$), and hydrogels with calcium acetate ($G_{Ca}$).
To prepare pure hydrogels ($G_{pure}$), gellan powder was dispersed in ultrapure water at room temperature. The dispersion was first heated up in the microwave until it became transparent and then, for rheological measurements, it was poured into petri dishes, at room temperature, to obtain a gel thickness of about 4-5 mm. $G_{pure}$ were investigated at a polysaccharide mass fraction of 2\%wt. Hydrogels containing calcium or sodium ions were obtained by mixing an aqueous solution of gellan with an aqueous solution of the corresponding salt and by following the same protocol described for the hydrogels without added salts. Hydrogels with added salts were also prepared at a polysaccharide mass fraction of 2\%wt. $G_{Na}$ was prepared from 27 mM sodium chloride aqueous solutions, while $G_{Ca}$ was obtained using calcium acetate aqueous solutions at a concentration of 2.5 mM.

\section{Rheological Measurements}
Rheological measurements were performed with a Rheometer MCR102, Anton Paar with a plate-plate geometry (diameter=49.97 mm) equipped with a Peltier system to control temperature, an evaporation blocker and an isolation hood to prevent solvent evaporation. Measurements were performed through a ReoCompass software. Storage ($G^{\prime}$) and loss ($G^{\prime\prime}$) moduli were measured at T=25$^{\circ}$C as a function of shear strain $\gamma$ at frequency f=1 Hz, in linear viscoelastic regime. Measurements were carried out $\sim$18 and $\sim$43 hours after preparation, in order to test aging effects and reproducibility.

\section{AFM Measurements}
Atomic force microscopy measurements were performed with a Dimension Icon (Bruker AXS) instrument. Images were acquired in air, at room temperature and under ambient conditions in Tapping mode, in order to protect the samples from damage. To maximize the image resolution, a RTESP (Rotated Tapping Etched Silicon) probe (Bruker, Germany) with a nominal radius of curvature R $\leq$ 8 nm was employed. For all the AFM measurements, samples have been deposited on freshly cleaved mica, incubated less than 1 minute, then rinsed with Milli-Q water and analyzed. Images have been corrected by levelling and background subtraction using Gwiddion 2.56 free software. A minimum of 100 height profiles of aggregates have been used for size histograms.

\section{All-atom molecular dynamics simulations}

\subsection{Gellan force field}
\begin{figure}[htbp]
\centering
\includegraphics[width=0.5\textwidth]{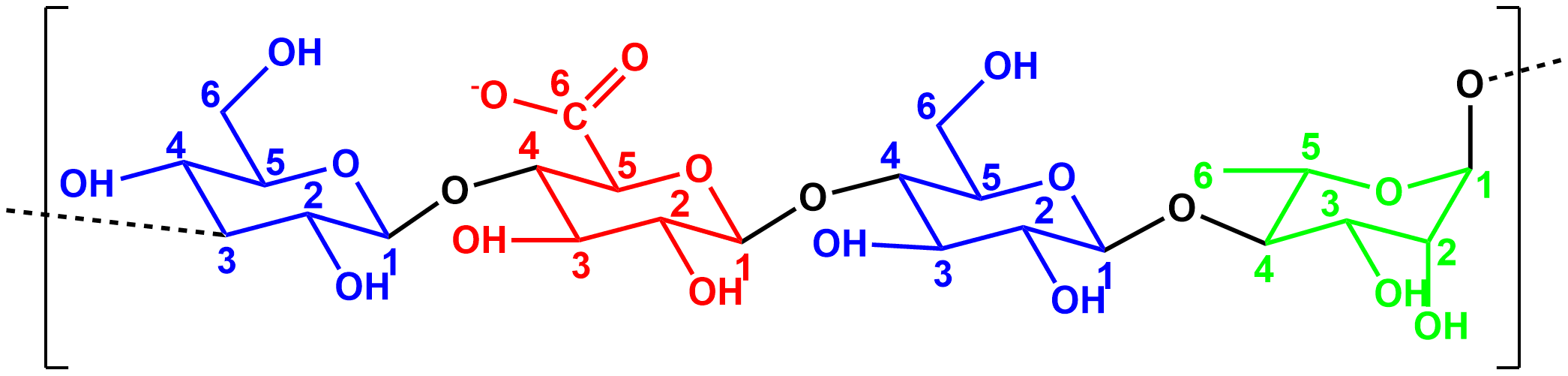}
  \caption{Chemical structure of gellan repeating unit which is composed by four monosaccharides: D-glucose, D-glucuronic acid, D-glucose and L-rhamnose that are shown in blue, red, blue, and green, respectively. D-glucuronic acid is represented in the ionized form. Glycosidic linkages are shown in black. Atomic numbering is displayed on carbon atoms.}
  \label{fgr:chem_struct}
\end{figure}

Starting atomic coordinates for the repeating unit of gellan were taken from the reported X-ray diffraction structure of oriented fibers of the gellan potassium salt~\cite{chandrasekaran1988cation}. Gellan force field was developed using the parameters values for the bonds, angles, dihedrals, and improper dihedrals based on the CHARMM36 force field~\cite{guvench2008additive} for carbohydrates and by calculating atomic partial charges. Charge calculation was performed on the gellan repeating unit having a methyl group bonded to the oxygen atoms of carbon 3 of D-glucose and carbon 1 of L-rhamnose (Figure~\ref{fgr:chem_struct}), in order to mimic the chain extension. To estimate electronic densities quantum mechanical calculations were carried out using the Gaussian16 software~\cite{g16}. The geometry of gellan was optimized at the B3LYP/6-31g(d) level of theory. In the calculation, a dielectric continuum was used to simulate the solvent effect, as previously reported~\cite{larwood1996solvation}. Single-point energy calculation on the geometry optimized gellan configuration was carried out at the HF/6-31g(d) level of theory. Effective charges were calculated by using the RESP method~\cite{bayly1993well}.

\subsection{Simulation protocol}
All-atom molecular dynamics simulations were performed on a dispersion of gellan chains in aqueous solution to characterize the mechanism of aggregation. A gellan chain consists of 4
repeating units, overall corresponding to 16 monosaccharides, and it is end-capped by methoxy groups. Gellan was modeled using the newly developed force field, except for a representative additional simulation of the system $G_{Ca}$ 5\%wt , that was carried out directly using CHARMM36~\cite{guvench2008additive} atomic partial charges as in Ref.~\cite{jamil2019dynamics}. This was done in order to validate the  force field developed in this work.
In all cases, salts and water were described using the CHARMM36~\cite{guvench2008additive} and TIP3P~\cite{jorgensen1983comparison}force fields, respectively. To investigate the effects of polysaccharide concentration and added salts, several experimental conditions were explored. In particular, simulations were carried out at gellan mass fraction of 3, 5 and 10\% without salts and also by adding sodium chloride or calcium acetate. Simulations with sodium chloride were performed with a salt concentration of 
0.1 M. Simulations with calcium acetate were carried out with a salt concentration of 
0.05 M, thus maintaining the same number of charges brought in solutions by the cations. For each simulated condition, first 6 energy optimized gellan chains with deprotonated glucuronic acid units were inserted in a cubic box and oriented in order to maximize the distance between each other. Then, the number of TIP3P water molecules corresponding to the set concentration, sodium counterions were added, and another energy minimization with tolerance of 100 $kJ$ $mol^{-1} nm^{-1}$ was carried out. The resulting system was equilibrated at 298 K for 40 ns and heated up to 353 K at 1 K $ns^{-1}$. For the simulations with added salts, a number of ions corresponding to the set concentration was added to the configuration at 353 K and the system was equilibrated for 40 ns at 353 K. Simulations were then carried out for 160 ns at 353 K. Finally the system was cooled to 298 K at 1 K $ns^{-1}$, equilibrated at 298 K for 40 ns and trajectory data were acquired for 160 ns at 298 K. This simulation protocol was chosen in order to mimic the preparation of gellan microgels put forward in Refs.~\cite{caggioni2007rheology,di2020gellan}.
All simulations were carried out in the NPT ensemble. The leapfrog integration algorithm~\cite{hockney1970potential} with a time step of 2 fs was used. Cubic periodic boundary conditions and minimum image convention were applied. The length of bonds involving H atoms was constrained by the LINCS procedure~\cite{hess1997lincs}. The velocity rescaling thermostat coupling algorithm, with a time constant of 0.1 ps was used to control temperature~\cite{bussi2007canonical}. During equilibration, pressure was maintained by using the Berendsen barostat~\cite{berendsen1984molecular} with a time constant of 1 ps. During data acquisition, pressure of 1 atm was maintained by the Parrinello-Rahman barostat, with a time constant of 2 ps~\cite{parrinello1981polymorphic,nose1983constant}. The cutoff of nonbonded interactions was set to 1 nm. Electrostatic interactions were calculated by the smooth particle-mesh Ewald method~\cite{essmann1995smooth}. Trajectories were acquired with the GROMACS software package (version 2018)~\cite{abraham2015gromacs,markidis2015solving} and the last 100 ns were considered for analysis, sampling 1 frame every 5 ps.

\section{Results and discussion}
\begin{figure}[htbp]
\centering
\includegraphics[width=0.5\textwidth]{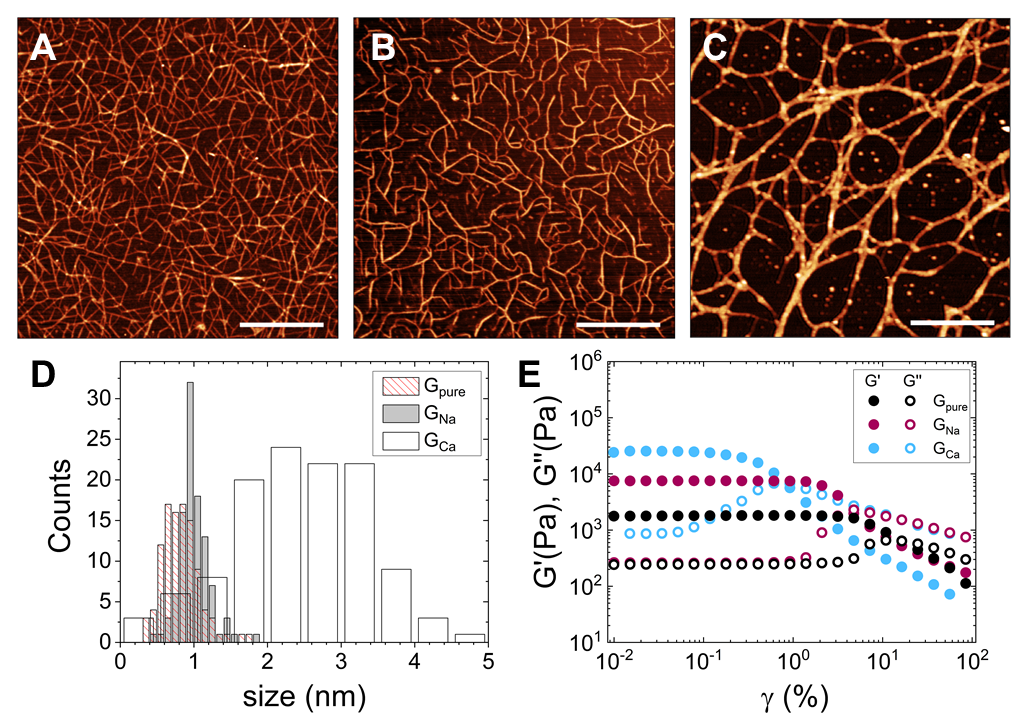}
  \caption{AFM images (bar = 500 nm) of gellan hydrogels with a polysaccharide concentration of 2\%wt (A) without added salts, (B) with sodium chloride 27 mM, and (C) with calcium acetate 2.5 mM. (D) Histograms of size distribution of gellan chains determined by height profile. (E) Storage $G^{\prime}$ (closed circles) and loss $G^{\prime\prime}$ (open circles) moduli as a function of shear strain $\gamma$ at f=1Hz and T=25$^{\circ}$C for gellan hydrogels at a polysaccharide concentration of 2\%wt with calcium acetate 2.5 mM (light blue) or sodium chloride 27 mM (bordeaux) compared with pure gellan (black).}
  \label{fgr:exp}
\end{figure}

The polysaccharide gellan is known to form physical gels through electrostatic interactions favoured by the presence of cations. In order to characterize the molecular mechanism of formation of gellan gels and to understand the role played by the cations, first we have carried out a qualitative investigation of the structure of gellan aggregates formed in different experimental conditions through Atomic Force Microscopy (AFM). Figures~\ref{fgr:exp}A-C directly compare the images of the polymer network obtained without adding salts to the polymer suspension ($G_{pure}$) and by adding sodium chloride ($G_{Na}$) or calcium acetate ($G_{Ca}$).
We focus on results for $G_{Na}$ (27 mM) and $G_{Ca}$ (2.5 mM), since these are the salt concentrations relevant for application to paper preservation for microgels~\cite{di2020gellan} and hydrogels~\cite{mazzuca2014gellan}, respectively.

While the networks of $G_{pure}$ and $G_{Na}$, Figures~\ref{fgr:exp}A-B respectively, show similar features, being composed by gellan aggregates of smaller size, aggregation in $G_{Ca}$ (Figure~\ref{fgr:exp}C) appears more pronounced, as revealed by the distribution of the transversal size of polysaccharide filaments (aggregates), which widens and shifts at higher values (Figure~\ref{fgr:exp}D). To determine the effect of the structural differences in the polymer network on the mechanical properties of the hydrogels, rheological measurements were performed on the same $G_{pure}$, $G_{Na}$, and $G_{Ca}$ samples. Figure~\ref{fgr:exp}E shows the storage and loss moduli $G^{\prime}$ and $G^{\prime\prime}$ as a function of the shear strain $\gamma$. In all conditions the polymer network behaves as an elastic solid with $G^{\prime}$ greater than $G^{\prime\prime}$. However, the addition of sodium causes a rise mostly of $G^{\prime}$, while the addition of calcium induces an increase of both $G^{\prime}$ and $G^{\prime\prime}$ revealing a different stability of the polymer network. Aging effects were also tested by measuring the samples at different times after the preparation and no significant differences were detected (Figure S1 of the Supporting Information). Figure~\ref{fgr:exp}E shows that the critical strain $\gamma_c$, which is defined as the intersection point between $G^{\prime}$ and $G^{\prime\prime}$ and corresponds to the breaking point of the gel, occurs at slightly lower shear stress $\gamma$ when sodium is added to the polysaccharide suspension. Moreover $\gamma_c$ is strongly reduced in the presence of calcium, suggesting an increase of stiffness and a significant different mechanical behavior of the gellan network formed with this divalent cation.

\begin{figure}[htbp]
\centering
\includegraphics[width=0.5\textwidth]{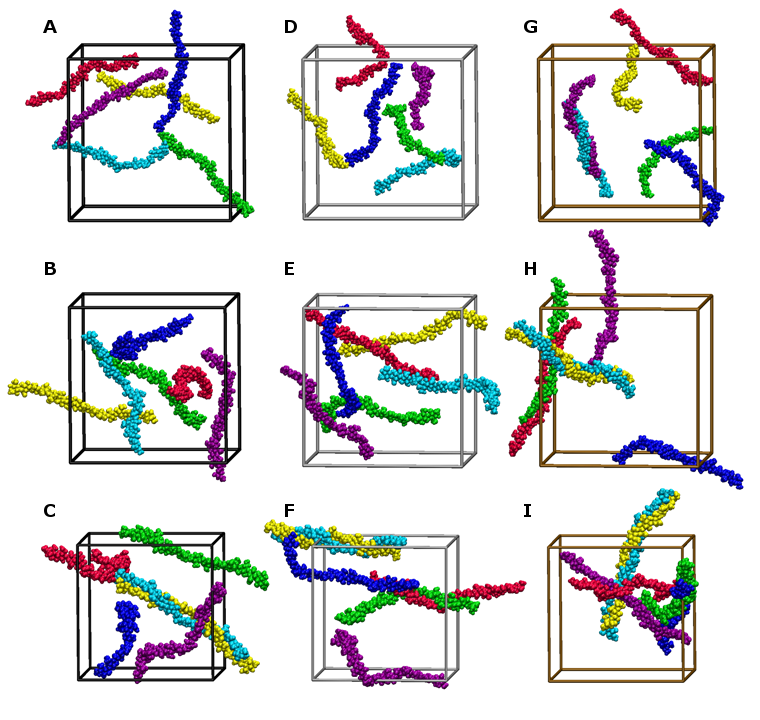}
  \caption{Representative snapshots from all-atom simulations at 298 K of pure gellan (black simulation box) at concentration of (A) 3, (B) 5, and (C) 10\%wt; gellan with sodium chloride 0.1 M (gray simulation box) at a polysaccharide concentration of (D) 3, (E) 5, and (F) 10\%wt; and gellan with calcium acetate 0.05 M (gold simulation box) at a polysaccharide concentration of (G) 3, (H) 5, and (I) 10\%wt. Each polysaccharide chain is shown with a different color. Ions and water molecules are omitted for clarity.}
  \label{fgr:snap}
\end{figure}

\begin{figure}[htbp]
\centering
\includegraphics[width=0.45\textwidth]{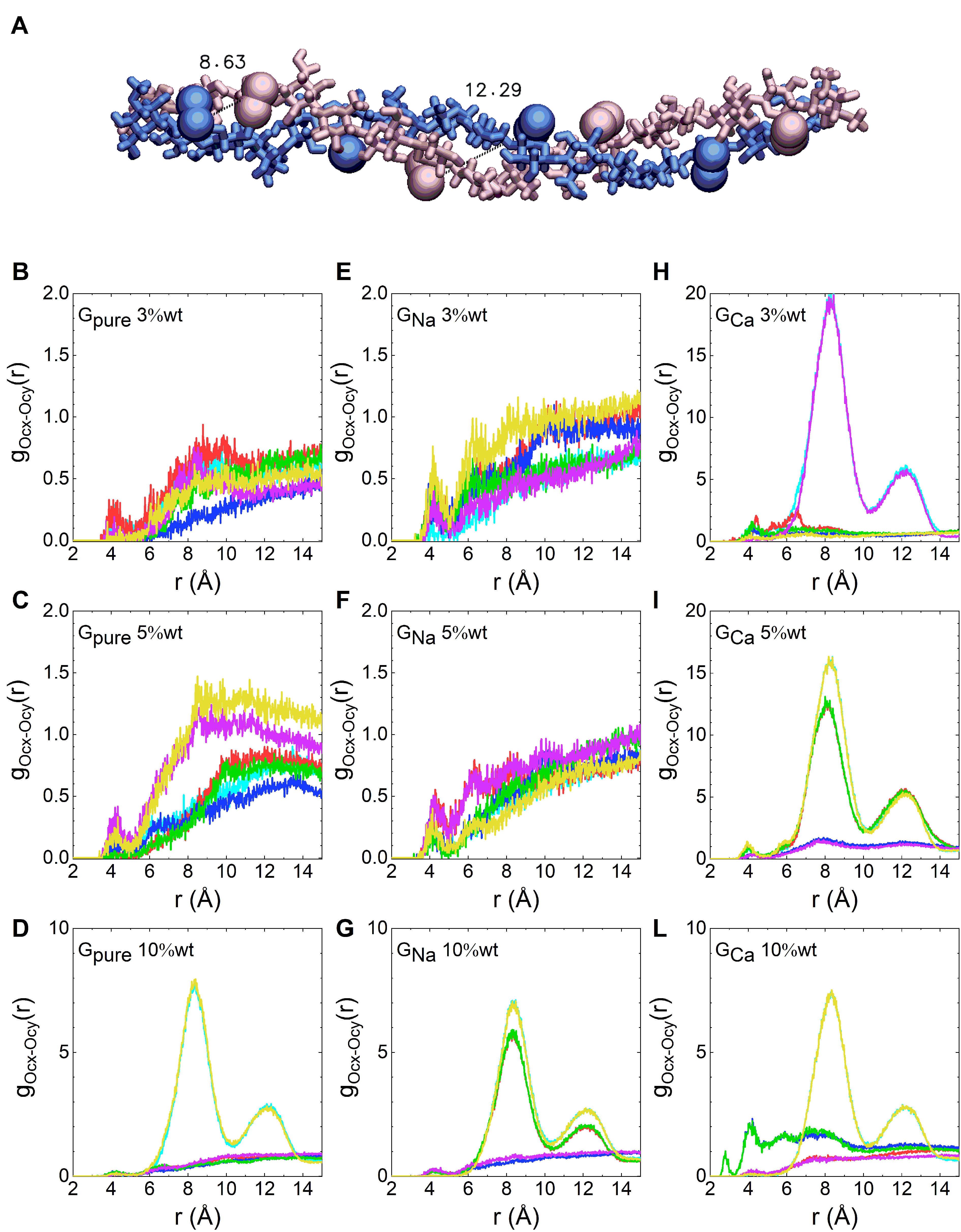}
  \caption{(A) Double helix structure formed from the association of two gellan chains in the system $G_{Ca}$ 5\%wt. Oxygen atoms belonging to the carboxylate group of the D-glucuronic acid unit and their characteristic interchain distances are highlighted in the representation. Radial distribution functions for gellan oxygen atoms of the carboxylate group of the chain ($C_x$) with respect to all other chains ($C_y$) calculated at 298 K for pure gellan at concentration of (B) 3\%wt, (C) 5\%wt, and (D)10\%wt; gellan with sodium chloride 0.1 M at a polysaccharide concentration of (E) 3\%wt, (F) 5\%wt, and (G)10\%wt; and gellan with calcium acetate 0.05 M at a polysaccharide concentration of (H) 3\%wt, (I) 5\%wt, and (L)10\%wt. Data calculated for chain 1, 2, 3, 4, 5, and 6 are shown in cyan, red, blue, green, purple, and yellow, respectively.}
  \label{fgr:rdfoo}
\end{figure}

To identify the molecular mechanism responsible for the differences in the macroscopic structural and mechanical features of gellan hydrogels, we exploited all-atoms molecular dynamics simulations. To this aim, we use the new gellan force field, as described in the Section All-atom molecular dynamics simulations methods, and performed numerical simulations at ambient temperature (298 K) of a suspension of gellan chains in water at a polysaccharide mass fraction of 3, 5 and 10\%wt. In addition to the suspension of gellan chains in pure water, which contains only the carboxylate counterions that are mostly sodium ions, we investigated the behavior of gellan chains in aqueous solution of salts, such as sodium chloride 0.1 M and calcium acetate 0.05 M. We studied salts concentrations bringing the same amount of positive charges in order to compare the effect of the chemical valence of the cation. Figures~\ref{fgr:snap}A-I display some representative configurations of the gellan chains simulated in the different experimental conditions. For the systems $G_{pure}$ and $G_{Na}$ aggregation of gellan chains is detected only at the highest concentration of 10\%wt. Differently, for $G_{Ca}$ association of gellan chains is observed even at the lower concentration of 3\%wt and it becomes more extended at higher polysaccharide concentrations. We note that, independently on the investigated aqueous solution conditions, in all our simulations association between pair of gellan chains occurs through the formation of double helix structures (Figure~\ref{fgr:rdfoo}A), in agreement with experiments~\cite{grinberg2003thermodynamics,takahashi2004solution}. Hence, differently from what determined by recent AFM experiments~\cite{diener2020rigid}, we did not detect the formation of gellan single helix. This may be attributed to the effect of the molar mass on the coil to helix transition, which requires a lower critical molar mass to favour the transition~\cite{ogawa2006effects} with respect to that used in the present simulations. This point would deserve further investigation in the future.

We have further inspected our proposed gellan force field, by comparing simulations results for the system $G_{Ca}$ 5\%wt, in which strong aggregation is experimentally observed, with those obtained from a supplementary simulation carried out by using CHARMM36~\cite{guvench2008additive} atomic partial charges in the same conditions (see Table S1 in the Supporting Information). While by using our newly developed force field, we detect pronounced aggregation of gellan (Figure~\ref{fgr:snap}H), in agreement with the experimental observations (Figure~\ref{fgr:exp}C), results from simulations carried out with CHARMM36~\cite{guvench2008additive} atomic partial charges neither reproduce the extended aggregation of the system, nor show the formation of helix structures. This is illustrated in Fig.~S2 of the Supporting Information, validating the gellan force field put forward in this work.

We characterized in more detail the aggregation of gellan chains in aqueous solution by evaluating the radial distribution functions of the gellan oxygen atoms belonging to the carboxylate group of the D-glucuronic acid unit (see the chemical structure of the repeating unit shown in Figure~\ref{fgr:chem_struct}). The results are summarized in Figures~\ref{fgr:rdfoo}B-L which display all the distribution functions calculated for each individual chain with respect to the others. For the systems $G_{pure}$ and $G_{Na}$ the radial distribution functions show well defined peaks at $\sim$8 and $\sim$12 {\AA} only at the concentration of 10\%wt. Instead in the case of $G_{Ca}$ at all concentrations two peaks are detected at the same distances. By analyzing the structure of the double helix formed by the association between two gellan chains (Figure~\ref{fgr:rdfoo}A), it is possible to assign the two peaks to the pairs of oxygen atoms each of them belonging to a different chain which, due to the conformation of the double helix, are pointing towards the same ($\sim$8 {\AA}) or opposite ($\sim$12 {\AA}) direction. The lowest height peaks that arise at lower distances mainly in $G_{Ca}$ can be attributed to less frequent contacts between gellan chains not associated in a double helix structure.

\begin{figure}[htbp]
\centering
\includegraphics[width=0.45\textwidth]{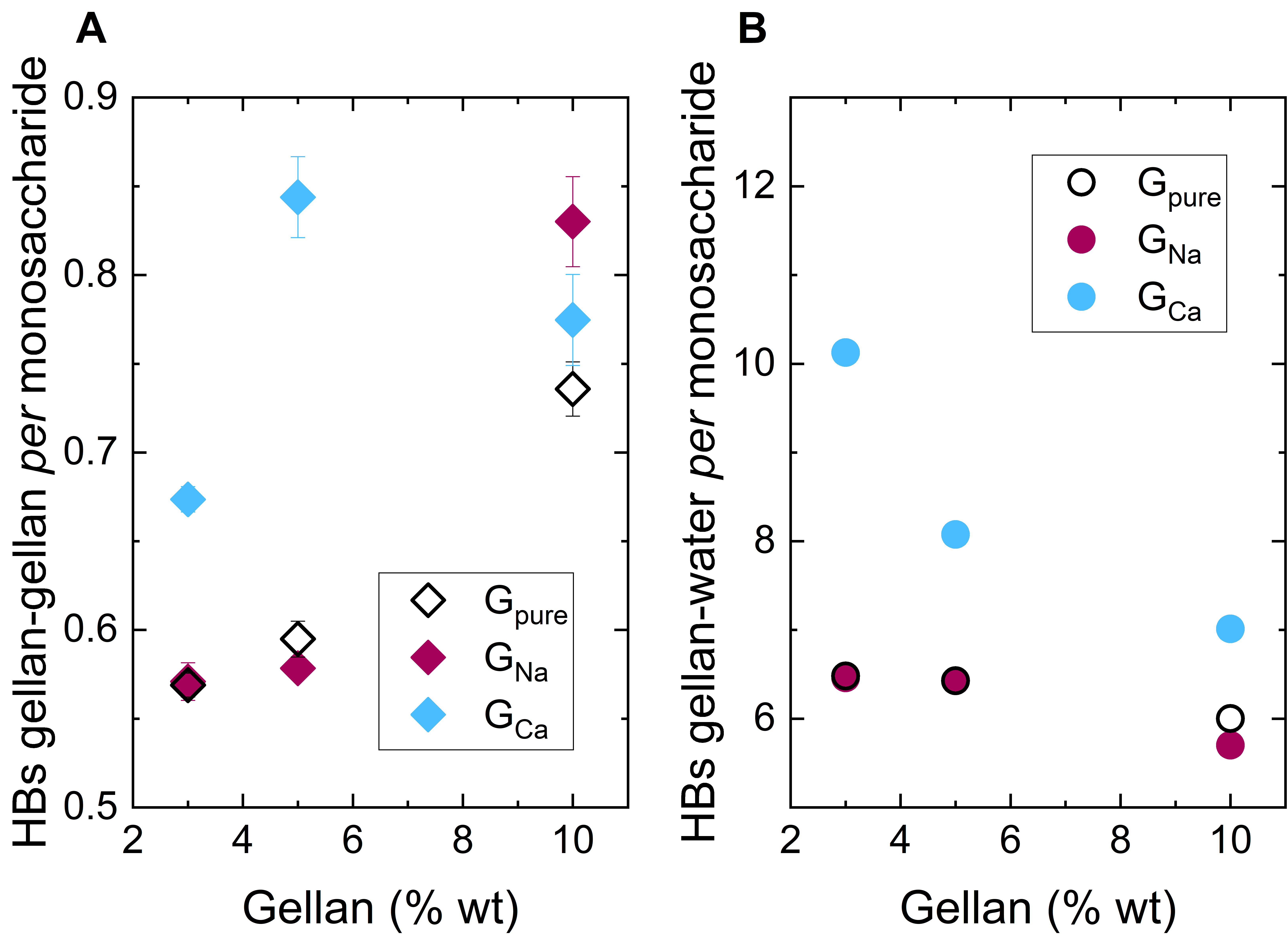}
  \caption{Average number of gellan-gellan (A) and gellan-water (B) hydrogen bonds normalized to the number of monosaccharides and averaged over the last 100 ns of simulations as a function of the gellan concentration. Data calculated for the simulations of gellan without added salts, with sodium chloride and with calcium acetate are shown in black, bordeaux, and light blue, respectively. Errors are estimated by the blocking method.}
  \label{fgr:totHB}
\end{figure}

We further investigated the solution structuring by monitoring the hydrogen bonding interactions formed by the gellan chains and those occurring between water and the polysaccharide, as reported in Figure~\ref{fgr:totHB}A and Figure~\ref{fgr:totHB}B. The average number of intra- and inter-chain hydrogen bonds formed by gellan is higher in the presence of calcium and, differently from $G_{pure}$ and $G_{Na}$ in which it increases with the polysaccharide content, a non-monotonic concentration dependence is observed in $G_{Ca}$ (Figure~\ref{fgr:totHB}A), in agreement with the experimental observation that high salt concentrations cause a decrease in gel strength~\cite{iurciuc2016gellan}. The number of hydrogen bonds between associated gellan chains appears to be stable, as shown by the time evolution of the gellan hydrogen bonds formed by each chain with all others reported in Figure S3 of the Supporting Information. In addition, the calcium cation not only favors the interactions between gellan chains, but it also increases the structuring of hydration water molecules, promoting a higher average number of water-gellan hydrogen bonds in the whole concentration range explored (Figure~\ref{fgr:totHB}B).

\begin{figure}[htbp]
\centering
\includegraphics[width=0.45\textwidth]{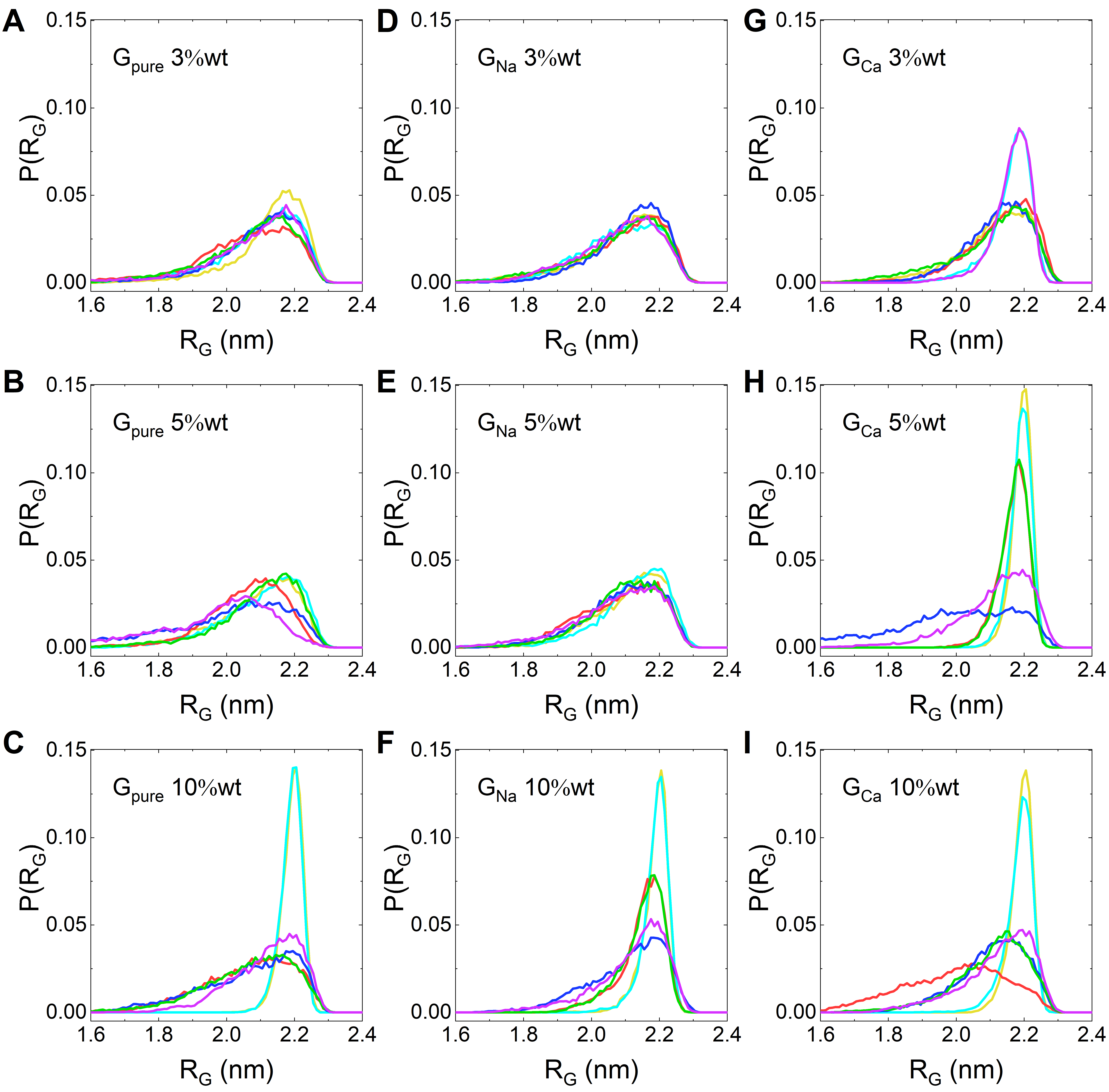}
  \caption{Distribution of the radius of gyration ($P(R_G)$) of each gellan chain at 298 K calculated for pure gellan at concentration of (A) 3\%wt, (B) 5\%wt, and (C)10\%wt; gellan with sodium chloride 0.1 M at a polysaccharide concentration of (D) 3\%wt, (E) 5\%wt, and (F)10\%wt; and gellan with calcium acetate 0.05 M at a polysaccharide concentration of (G) 3\%wt, (H) 5\%wt, and (I)10\%wt. Data calculated for chain 1, 2, 3, 4, 5, and 6 are shown in cyan, red, blue, green, purple, and yellow, respectively.}
  \label{fgr:prg}
\end{figure}

We now examine the conformation of the polysaccharide in aqueous solution. Figures~\ref{fgr:prg}A-I report the distribution of radius of gyration of each individual chain for the different conditions explored. In the systems where no aggregation is detected, such as $G_{pure}$ 3\%wt (Figure~\ref{fgr:prg}A) or $G_{Na}$ 3\%wt (Figure~\ref{fgr:prg}D), all gellan chains assume a similar broad distribution of radius of gyration. Conversely, when association occurs, the conformation of gellan chains involved in the formation of aggregates can be clearly distinguished from the others. For example, in the system $G_{Ca}$ 5\%wt (Figure~\ref{fgr:prg}H), the polysaccharide chains that are associated to other chains, as determined in Figure~\ref{fgr:snap}H and Figure~\ref{fgr:rdfoo}I, have a narrow distribution of radius of gyration with a maximum at $\sim$2.2 nm. Moreover, chains that form a partial double helix structure (chains red and green in Figure~\ref{fgr:prg}H) are characterized by a broader distribution of radius of gyration with respect to those fully paired (cyan and yellow in Figure~\ref{fgr:prg}H). These findings suggest that gellan aggregation occurs through a conformational transition of the polysaccharide chains which leads to rigid elongated conformations. The lower flexibility of associated gellan chains is also detectable by monitoring the time evolution of the radius of gyration of each polysaccharide chain, as displayed in Figure S4 of the Supporting Information. For example, in the system $G_{Ca}$ 5\%wt (Fig. S4H) the fluctuations of the values of the radius of gyration for associated chains (chains cyan and yellow) are considerably reduced as compared to the other chains (chains purple and blue). To understand in more detail the molecular origin of the flexibility of gellan chains, we have further analyzed the behavior of the dihedral angles of the glycosidic linkages between the monosaccharides composing the chains. Because of the hindered torsion of sugar ring bonds, the rotation around the glycosidic linkage is the mechanism which confers to the polysaccharide partial flexibility. We have thus evaluated the number of transitions of all the glycosidic dihedral angles of each gellan chain $\Phi$, defined as H1-C1-O1-C4', and $\Psi$, defined as C1-O1-C4'-H4', as illustrated in the Scheme S1 and shown in Figures S5A-I and S6A-I of the Supporting Information. The results indicate that the rotation of the dihedral angle $\Phi$ is quite restricted, independently on the gellan concentration, added salt, or aggregation state of the chain. In the case of the dihedral angle $\Psi$, we observe that (i) transitions are mostly detected for glycosidic linkages to glucose residues; (ii) no direct correlation with the nature of the cation added is observed and (iii) the dihedrals transitions are hindered for associated gellan chains, determining the increased stiffness of aggregated systems.

\begin{figure}[htbp]
\centering
\includegraphics[width=0.5\textwidth]{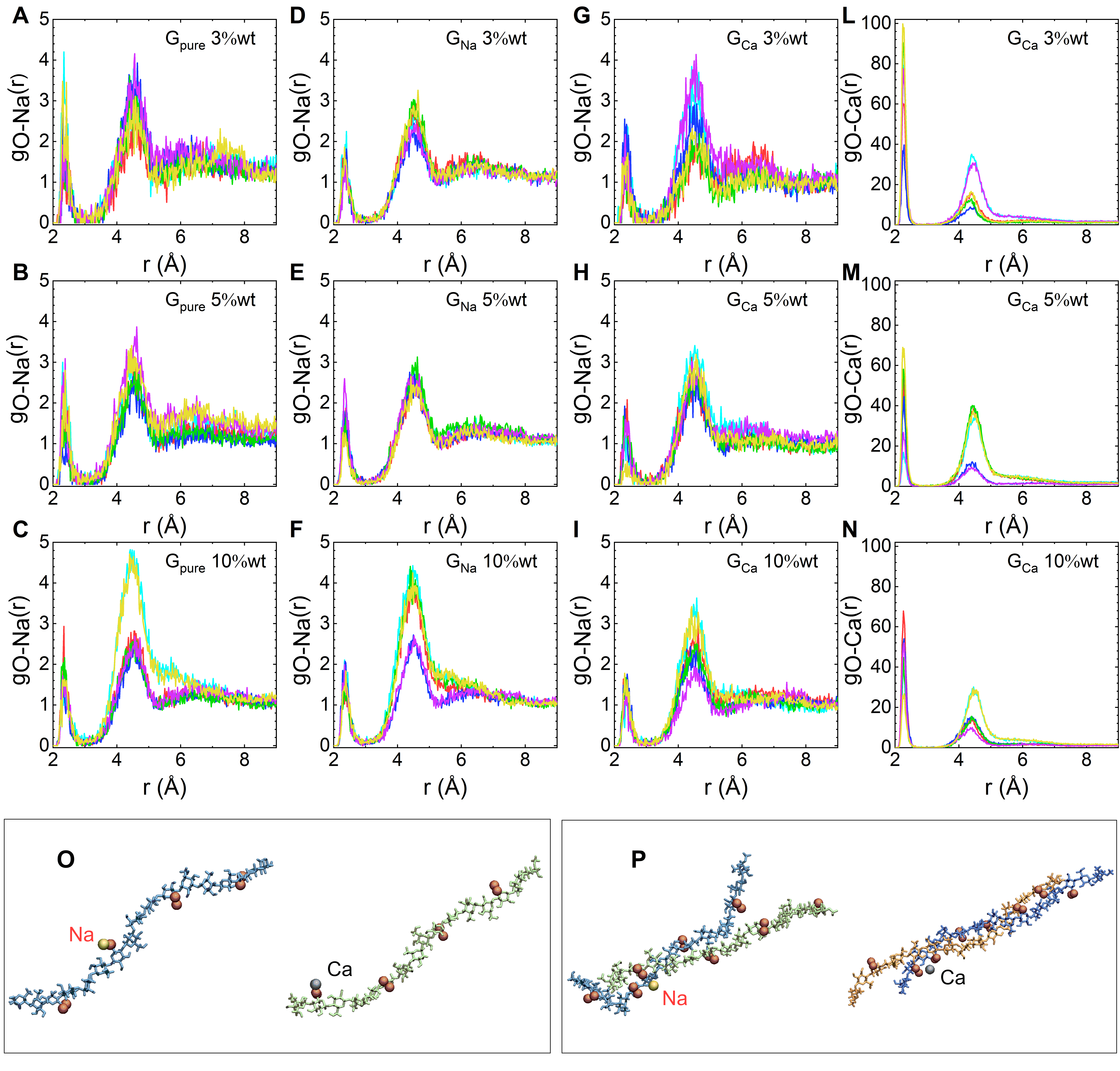}
  \caption{Radial distribution functions for sodium cations around the gellan oxygen atoms of the carboxylate group of each chain calculated at 298 K for pure gellan at concentration of (A) 3\%wt, (B) 5\%wt, and (C)10\%wt; gellan with sodium chloride 0.1 M at a polysaccharide concentration of (D) 3\%wt, (E) 5\%wt, and (F)10\%wt; and gellan with calcium acetate 0.05 M at a polysaccharide concentration of (G) 3\%wt, (H) 5\%wt, and (I)10\%wt. Radial distribution functions for calcium cations around the gellan oxygen atoms of the carboxylate group of each chain calculated at 298 K for gellan with calcium acetate 0.05 M at a polysaccharide concentration of (L) 3\%wt, (M) 5\%wt, and (N)10\%wt. Data calculated for chain 1, 2, 3, 4, 5, and 6 are shown in cyan, red, blue, green, purple, and yellow, respectively. Representative snapshots showing the arrangement of calcium and sodium around gellan carboxylate groups of a single chain (O) or a double helix (P). Sodium and calcium are shown in yellow and grey, respectively. Oxygen atoms of the gellan carboxylate groups are displayed in ochre.}
  \label{fgr:rdfoocat}
\end{figure}

Thanks to the present simulation results, we now tackle the two main open questions about gellan aggregation mechanism. First, we address the issue of helix formation. The results reported so far clearly illustrate that double helices form in the presence of calcium at all studied gellan concentrations. They also form in the presence of added sodium or in pure gellan, but only for the most concentrated polymer conditions. Hence, it appears that bivalent cations definitely facilitate the occurrence of the double helix, but their presence is not a necessary condition. In this respect, it is useful to recall that even in pure gellan an amount of dissolved electrolyte is present. For example in our simulations each polysaccharide chain adds to the solution 4 sodium counterions for a total of 24 ions in every simulated system. As a comparison, the gellan suspension at concentration of 5\%wt in sodium chloride 0.1 M contains 58 sodium ions, 24 of them being the gellan counterions. Hence, at high enough gellan concentration, gellan counterions themselves will have an effect on the aggregation process. We now want to discriminate the specific role of the cations by assessing what is their contribution to the double helix formation with a closer look to how cations are arranged with respect to chain. We start by focusing on the carboxylate groups because we expect these to be the ones undergoing more significant interactions with the cations due to their opposite charge and report the radial distribution functions for sodium or calcium with respect to the gellan oxygen atoms of these groups $g_{O-Na/Ca}(r)$ in
Figures~\ref{fgr:rdfoocat}A-N for all studied state points. In particular, each plot reports six curves corresponding to each individual gellan chain in the simulations, so that we can discriminate the behavior of those chains involved in double helix formation.
For all cases, we observe the presence of two peaks in $g_{O-Na/Ca}(r)$ at short distances, roughly located at the same two distances for all chains, independently of their aggregation state. In particular, a first peak occurs around 2.3 {\AA}, corresponding to cations directly interacting with the carboxylate oxygens (snapshots in Figure~\ref{fgr:rdfoocat}O), followed by a second peak located around 4.5 {\AA}. The latter distance is compatible either with cations being on opposite side of the chains with respect to the considered oxygen atoms or with those "bridging" two chains forming a helix (snapshots in Figure~\ref{fgr:rdfoocat}P).  We note that the intensity of both peaks is approximately 10 times larger in the case of calcium with respect to sodium.
Focusing on chains exhibiting double helix formation, we find an enhancement of the second peak in the presence of sodium ions (Figures~\ref{fgr:rdfoocat}C and F), while a huge variation is detected for calcium (Figures~\ref{fgr:rdfoocat}L-N) that is also accompanied by a significant growth of the first peak.
These results prove, at the microscopic level, that the formation of
the double helix can occur both in the presence of sodium and calcium ions. Since the behavior of $g_{O-Na}(r)$ at high polysaccharide concentration (10\%wt) is almost identical for pure gellan  and for gellan with added NaCl (Figures~\ref{fgr:rdfoocat}C and F, respectively), we deduce that double helix formation takes place even in the absence of added salt, but always through the mediation of the cations. This is signalled by the increase of the second peak of  the radial distribution functions for helix-forming chains, which is compatible with the appearance of configurations where the cations are in between the two chains (see snapshots in Figure~\ref{fgr:rdfoocat}P). Instead, the huge change observed in the calcium arrangement around helix-forming chains demonstrates the dominant role of bivalent ions, which trigger the helix formation in our simulations at all studied gellan concentrations.
Given that for sodium the same is true only at very large gellan concentrations (or presumably at high sodium concentrations, as suggested by experiments~\cite{robinson1991conformation}), the present results clearly show that monovalent salt ions only mediate helix formation as a consequence of their high density. {\it De facto} they seem just to participate to the phenomenon in crowded conditions, but cannot be considered primary actors for its occurrence as bivalent cations.

\begin{figure}[htbp]
\centering
\includegraphics[width=0.5\textwidth]{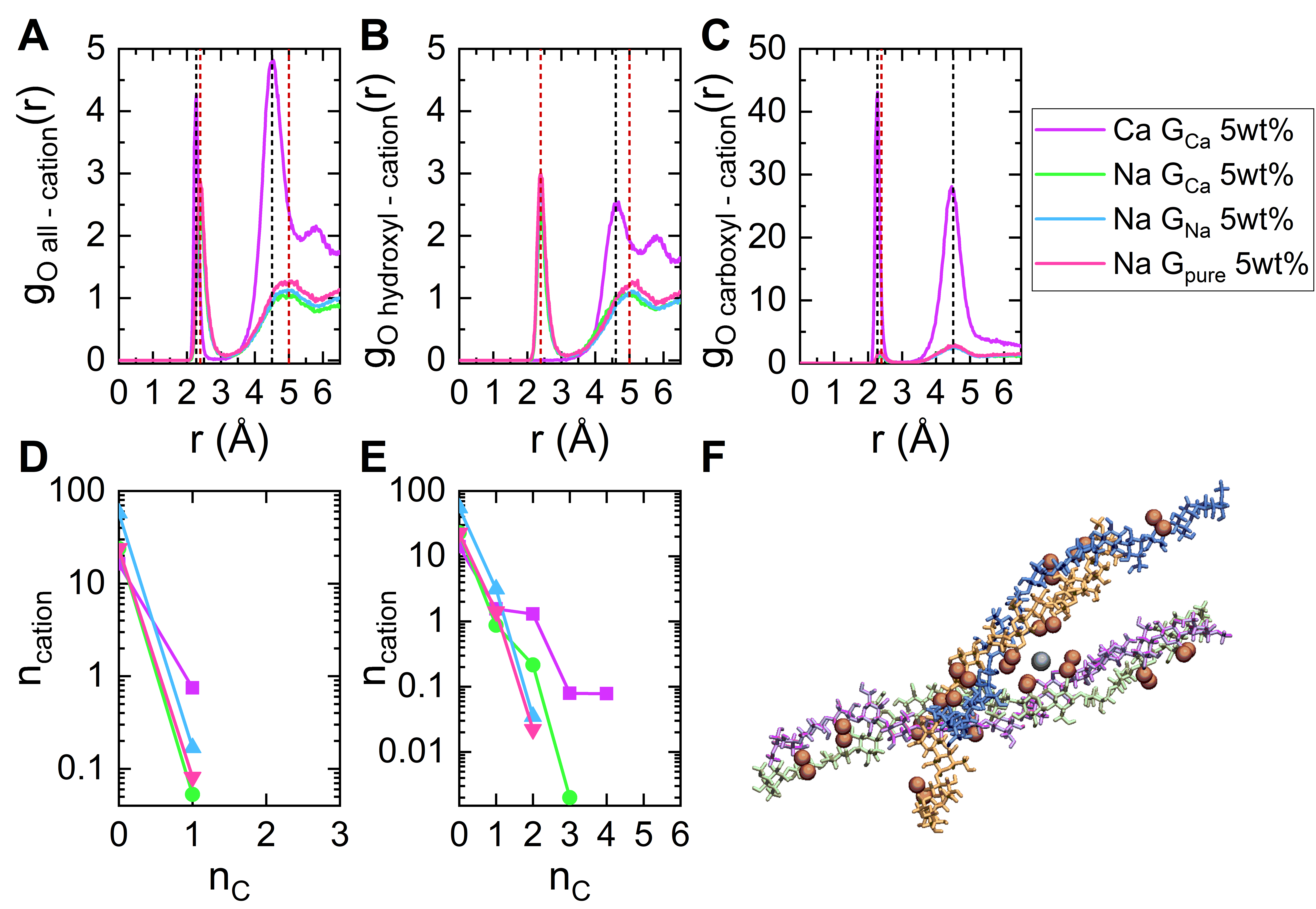}
  \caption{Radial distribution functions for cations around (A) all gellan oxygen atoms, (B) gellan oxygen atoms of the hydroxyl groups, and (C) gellan oxygen atoms of the carboxylate group. Vertical dashed lines mark the position of the peaks for sodium (red) and calcium (black). Distribution of the number of cations ($n_{cation})$ having $n_c$ neighbouring chains calculated for a distance (D) 0 {\AA} $\le d \le$ 4 {\AA} and (E) 4 {\AA} $\le d \le$ 6 {\AA} averaged over 100 ns of simulation. Data refers to the systems with gellan concentration of 5\%wt for pure gellan (pink), gellan with sodium chloride 0.1 M (blue), and gellan with calcium acetate 0.05 M for sodium (green) and calcium (purple). (F) Representative snapshot showing a super-aggregate between two double helix structures bridged by calcium, which is shown in grey. Oxygen atoms of the gellan carboxylate groups are displayed in ochre.}
  \label{fgr:rdfall}
\end{figure}

Having shed light on the double helix formation, we now turn to the `super-aggregates'. Of course, we are aware that the data collected in our simulations are limited in size, time and number of examined state points, but we still believe they are able to provide a convincing evidence of the crucial role of bivalent ions in this second step. Indeed, only for calcium acetate we have observed, within our numerical accuracy, the onset of the bonding of two double helices, coherently with the increased aggregate size measured by AFM (Figures~\ref{fgr:exp}D). Since this occurs only for the specific gellan concentration 5\% wt, we now focus on this case and  investigate the radial distribution function of the different cations with respect to the polysaccharide chains in  Fig.~\ref{fgr:rdfall}. This time we do not discriminate among different chains and we thus average results over all chains, but compare the findings for $gOall-cation(r)$, sampled for sodium or calcium cations close to all gellan oxygen atoms in Fig.~\ref{fgr:rdfall}A with the radial distributions calculated between gellan hydroxyl groups (Figure~\ref{fgr:rdfall}B) or  carboxylate groups (Figure~\ref{fgr:rdfall}C) and the cations. Comparing the three panels for sodium ions, we find that their average arrangement is always very similar, independently of the considered group, signaling a uniform distribution of these in the system, even in the presence of added calcium. On the contrary, the results for calcium ions are very interesting: for hydroxyl groups they are depleted at small distances, so that the close-contact peak is completely absent in this case; for carboxylate groups their signal is again 10 times larger and a long-range correlation also exists, for the high intensity of peaks detected at distances over 4 {\AA}.

Since the positions of the two first peaks are the same as those discussed in Figures~\ref{fgr:rdfoocat}A-N, the characteristic cation arrangement must belong to the cases described before. Since we cannot rely on the distance to reveal the presence of the super-aggregates, we therefore count how many chains are close to each cation within our simulation trajectories. The resulting distributions $n_{cation}$, denoting the number of cations having $n_C$ neighbouring chains, are reported in Fig.~\ref{fgr:rdfall}D and Fig.~\ref{fgr:rdfall}E, respectively for the two characteristic regions: 0-4 {\AA} for those in direct contact and 4-6 {\AA} for those mediating double helices and super-aggregates.
We observe that for the first shell, as expected, no cation exceeds one neighbouring chain, with calcium ions being preferentially closer to them.
However, in the second shell, a few sodium ions close to two chains appear, although from the analysis presented earlier, no double helices are formed at this gellan concentration. We note that the cases of pure gellan and gellan with sodium are very similar to each in all respects, even though the number of sodium ions in the system increases from 24 to 58 from one to the other condition. On the other hand, sodium ions in the presence of calcium have a different behavior, since considerably more ions are detected close to two chains as well as a rare event of three-chain connection. The most significant
result is found for calcium ions, despite they are the lowest in number (only 17), they show the largest number close to two chains, but also close to 3 and 4 chains. The latter provides unambiguous evidence that calcium ions promote super-aggregate formation, even in our very reduced and idealized simulation environment, as shown in the snapshot reported in Fig.~\ref{fgr:rdfall}E.

These results allow us to shed light on the second step mechanism of gellan gelation. Indeed, they suggest that bivalent salt clearly promotes in a rapid (because it happens within the duration of our simulations) and efficient (because despite their low number all our samples show enhanced gellan aggregation in their presence) aggregation of gellan chains. This is in agreement with the proposed interpretation of Gunning et al.~\cite{gunning1990light}, where a highly coherent super-structure and degree of cross-linking should be formed, as a sort of `strong gels'. A recent AFM investigation has also reported the formation of thick gellan fibrillar structures by lateral aggregation in the presence of calcium, confirming the extended aggregation promoted by divalent cations~\cite{diener2020rigid}. However, sodium ions, both as gellan counterions and as added electrolyte, surely participate to the mechanism. In the presence of calcium their rearrangement close to gellan chains is facilitated, but when they are on their own they are highly inefficient. Only at very large concentrations, they have an effect and it is thus likely that gels obtained in the presence of monovalent salt, are much more disordered, with an overall lower degree of crosslinking and likely the presence of disordered coils not assembled into double helices, giving rise to weaker gels, in agreement with the hypothesis of Robinson et al~\cite{robinson1991conformation}.

\section{Conclusions}
In this work we developed an atomistic model for the polysaccharide gellan to probe its gelation mechanism at the molecular scale. We performed extensive molecular dynamics simulations as a function of the polysaccharide concentration and with the addition of monovalent or divalent salts, in order to understand the steps involved in the aggregation process and to address the role played by cations. Our findings clearly show that gellan aggregation is a two-step process that is strongly affected by the nature of the cations involved. In the first step, a disorder-order transition from polysaccharide coils to double helix is observed. Then, in the second step, aggregation of double helices into super-structures occurs. However, our results point out at different roles played by monovalent and divalent cations. In the case of calcium, gellan aggregation is enhanced for both steps, as also supported by rheology and atomic force microscopy measurements. Indeed calcium cations favour the formation of double helices by complexation of the carboxylate groups belonging to different chains. Moreover, larger aggregates formed through calcium-mediated complexation of different double helices are detected. Differently, monovalent cations show a reduced effect on the process, allowing the transition to double helices only at very high salt and gellan content. Overall, the gelation mechanism observed in presence of divalent cations supports the interpretation proposed by Gunning et al.~\cite{gunning1990light}, where a highly ordered super-structure compatible with stronger gels is predicted. Instead, for monovalent cations only at large concentrations a net effect is detected and the resulting network structures show a lower degree of cross-linking, giving rise to weaker gels, as predicted by Robinson et al~\cite{robinson1991conformation}.

The present investigation can be further applied to study the aggregation mechanism of acylated and low-acylated gellan, in which the presence of acyl substituents increases the complexity of the aggregation process by conferring to the polysaccharide hydrophobic properties that can also lead to hydrophobic association and could inhibit or even preempt the formation of double helices. In this perspective, the use of atomistic simulations could be again crucial to discern between molecular mechanisms of aggregation that can generate different structures-properties functions of the material, thus allowing to identify the optimal conditions with respect to the specific application requirements.

\begin{acknowledgements}
We thank Lorenzo Gontrani for his help in performing quantum mechanical calculations. We acknowledge financial support from Regione Lazio (through L.R. 13/08 Progetto Gruppo di Ricerca MICROARTE n. prot. A0375-2020-36515).
We also acknowldge CINECA-ISCRA for HPC resources and CNIS - Research Center for Nanotechnology applied to Engineering of Sapienza, Sapienza University, Rome for AFM instrument.
\end{acknowledgements}

\renewcommand{\thefigure}{S\arabic{figure}}
\renewcommand{\thetable}{S\arabic{table}}
\setcounter{figure}{0}

\section{Supporting Information}

\subsection{S1. Gellan hydrogels aging}
\begin{figure}[htbp]
\centering
\includegraphics[width=0.45\textwidth]{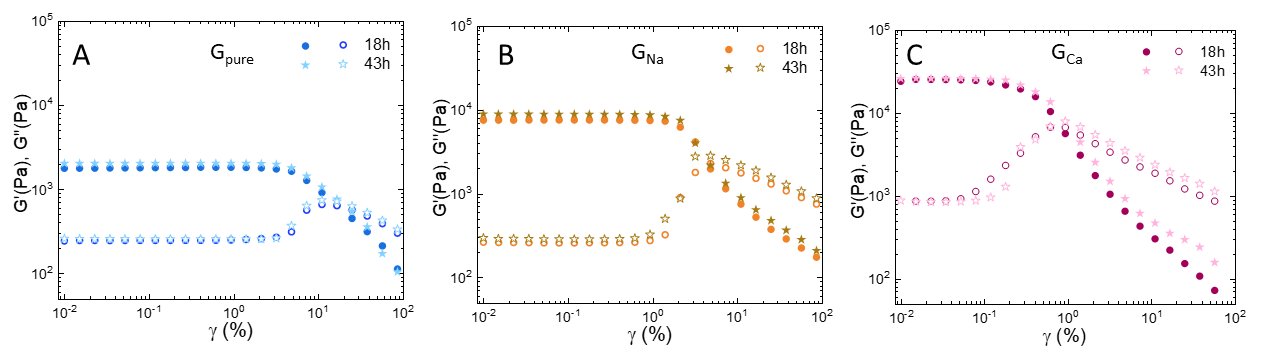}
  \caption{Storage and loss moduli $G^{\prime}$ and $G^{\prime\prime}$ as a function of the shear strain measured at different times after preparation (18 h and 43 h) for (A) $G_{pure}$, (B)  $G_{Na}$, and (C) $G_{Ca}$ samples.}
  \label{fgr:aging}
\end{figure}

To monitor aging effects, rheological measurements were performed on the same $G_{pure}$, $G_{Na}$, and $G_{Ca}$ samples at different times after the preparation. Figures~\ref{fgr:aging}A-C report the comparison of the storage and loss moduli $G^{\prime}$ and $G^{\prime\prime}$ as a function of the shear strain $\gamma$ at two different times showing that no significant changes can be detected.

\subsection{S2. Gellan force field}
\begin{figure}[htbp]
\centering
\includegraphics[width=0.4\textwidth]{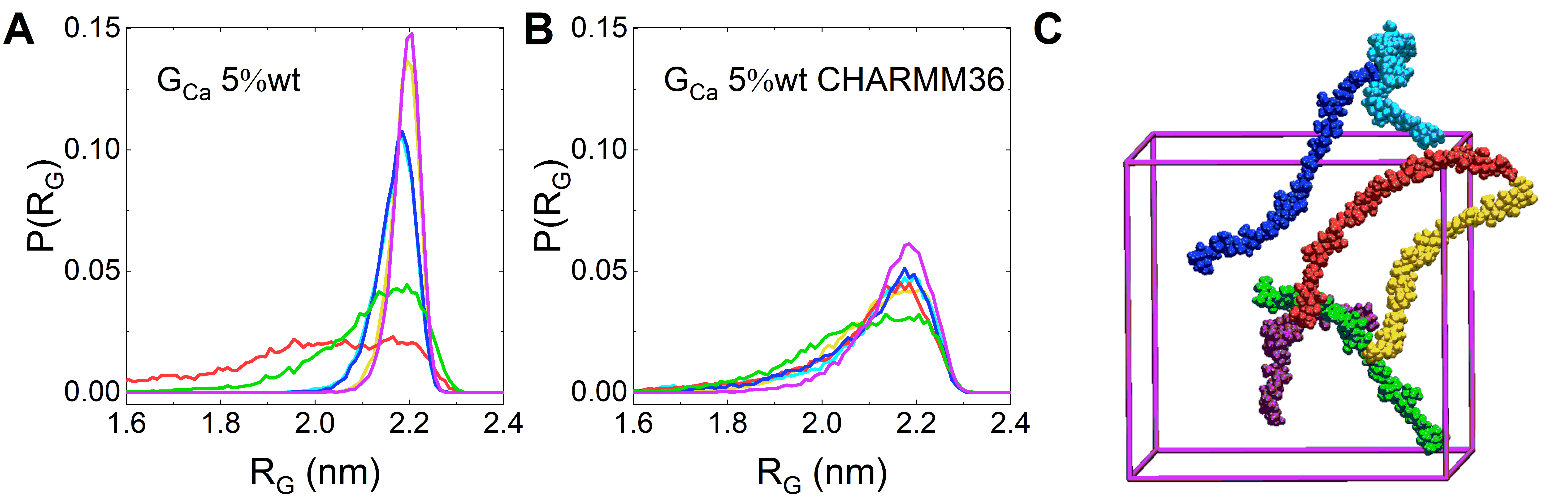}
  \caption{Distribution of the radius of gyration ($P(R_G)$) of each gellan chain at 298 K calculated for gellan with calcium acetate 0.05 M at a polysaccharide concentration of 5\%wt from the simulations performed using (A) the newly developed and (B) CHARMM36 atomic partial charges. Data calculated for chain 1, 2, 3, 4, 5, and 6 are shown in cyan, red, blue, green, purple, and yellow, respectively. (C) Representative snapshot from all-atom simulations with CHARMM36 atomic partial charges at 298 K of gellan with calcium acetate 0.05 M at a polysaccharide concentration of 5\%wt. Each polysaccharide chain is shown with a different color. Ions and water molecules are omitted for clarity.}
  \label{fgr:charmm}
\end{figure}

We have evaluated the soundness of our force field to reproduce the experimental aggregation of gellan by comparing simulations results for gellan with calcium acetate 0.05 M at a polysaccharide concentration of 5\%wt obtained using the atomic partial charges of our newly developed force field with those reported in the CHARMM36 force field. The values of the atomic partial charges for the two force fields are summarized in Table~\ref{tbl:q}. Figures~\ref{fgr:charmm}A and ~\ref{fgr:charmm}B compare the distribution of the radius of gyration ($P(R_G)$) of each gellan chain in the two systems under comparison. While for our newly developed force field different broad and sharp distributions of the radius of gyration, attributable to different chain conformations, are observed (Figure~\ref{fgr:charmm}A), for CHARMM36 no variations are detected (Figure~\ref{fgr:charmm}B). Moreover, by using the atomic partial charges of CHARMM36 force field there is no formation of helix structures nor extended aggregation, as shown in the representative snapshot reported in Figure~\ref{fgr:charmm}C.

\begin{longtable}{cccc}
    \hline
    Group & Atom & CHARMM36 & new \\
    \hline
    \scriptsize
    $\beta$-D-glucose & C1  &  0.290 &  0.298667 \\
    $\beta$-D-glucose & C2  &  0.140 &  0.066474 \\
    $\beta$-D-glucose & C3  &  0.140 &  0.119218 \\
    $\beta$-D-glucose & C4  &  0.090 &  0.200254 \\
    $\beta$-D-glucose & C5  &  0.110 &  0.229691 \\
    $\beta$-D-glucose & C6  &  0.050 &  0.203684 \\
    $\beta$-D-glucose & O2  & -0.650 & -0.621853 \\
    $\beta$-D-glucose & O3  & -0.360 & -0.437022 \\
    $\beta$-D-glucose & O4  & -0.650 & -0.715397 \\
    $\beta$-D-glucose & O5  & -0.400 & -0.546499 \\
    $\beta$-D-glucose & O6  & -0.650 & -0.704953 \\
    $\beta$-D-glucose & H1  &  0.090 &  0.095651 \\
    $\beta$-D-glucose & H2  &  0.090 &  0.099043 \\
    $\beta$-D-glucose & H3  &  0.090 &  0.099146 \\
    $\beta$-D-glucose & H4  &  0.090 &  0.058444 \\
    $\beta$-D-glucose & H5  &  0.090 &  0.036496 \\
    $\beta$-D-glucose & H61 &  0.090 &  0.054588 \\
    $\beta$-D-glucose & H62 &  0.090 &  0.054588 \\
    $\beta$-D-glucose & HO2 &  0.420 &  0.402987 \\
    $\beta$-D-glucose & HO4 &  0.420 &  0.439005 \\
    $\beta$-D-glucose & HO6 &  0.420 &  0.450311 \\
    $\beta$-D-glucuronic acid & C1  &  0.290 &  0.257354 \\
    $\beta$-D-glucuronic acid & C2  &  0.140 &  0.318595 \\
    $\beta$-D-glucuronic acid & C3  &  0.140 &  0.042422 \\
    $\beta$-D-glucuronic acid & C4  &  0.090 &  0.087925 \\
    $\beta$-D-glucuronic acid & C5  &  0.110 &  0.042531 \\
    $\beta$-D-glucuronic acid & C6  &  0.520 &  0.875354 \\
    $\beta$-D-glucuronic acid & O2  & -0.650 & -0.673471 \\
    $\beta$-D-glucuronic acid & O3  & -0.650 & -0.718937 \\
    $\beta$-D-glucuronic acid & O4  & -0.360 & -0.395148 \\
    $\beta$-D-glucuronic acid & O61 & -0.760 & -0.825435 \\
    $\beta$-D-glucuronic acid & O62 & -0.760 & -0.825435 \\
    $\beta$-D-glucuronic acid & H1  &  0.090 &  0.094327 \\
    $\beta$-D-glucuronic acid & H2  &  0.090 &  0.112573 \\
    $\beta$-D-glucuronic acid & H3  &  0.090 &  0.083692 \\
    $\beta$-D-glucuronic acid & H4  &  0.090 &  0.105353 \\
    $\beta$-D-glucuronic acid & H5  &  0.090 &  0.066513 \\
    $\beta$-D-glucuronic acid & HO2 &  0.420 &  0.444322 \\
    $\beta$-D-glucuronic acid & HO3 &  0.420 &  0.480174 \\
    $\beta$-D-glucose & C1  &  0.290 &  0.201172 \\
    $\beta$-D-glucose & C2  &  0.140 &  0.023633 \\
    $\beta$-D-glucose & C3  &  0.140 &  0.251145 \\
    $\beta$-D-glucose & C4  &  0.090 &  0.205766 \\
    $\beta$-D-glucose & C5  &  0.110 & -0.035713 \\
    $\beta$-D-glucose & C6  &  0.050 &  0.213947 \\
    $\beta$-D-glucose & O2  & -0.650 & -0.623962 \\
    $\beta$-D-glucose & O3  & -0.650 & -0.670336 \\
    $\beta$-D-glucose & O4  & -0.360 & -0.488503 \\
    $\beta$-D-glucose & O5  & -0.400 & -0.331659 \\
    $\beta$-D-glucose & O6  & -0.650 & -0.695593 \\
    $\beta$-D-glucose & H1  &  0.090 &  0.110912 \\
    $\beta$-D-glucose & H2  &  0.090 &  0.116235 \\
    $\beta$-D-glucose & H3  &  0.090 &  0.106824 \\
    $\beta$-D-glucose & H4  &  0.090 &  0.068972 \\
    $\beta$-D-glucose & H5  &  0.090 &  0.095652 \\
    $\beta$-D-glucose & H61 &  0.090 &  0.045148 \\
    $\beta$-D-glucose & H62 &  0.090 &  0.045148 \\
    $\beta$-D-glucose & HO2 &  0.420 &  0.437287 \\
    $\beta$-D-glucose & HO3 &  0.420 &  0.437362 \\
    $\beta$-D-glucose & HO6 &  0.420 &  0.429254 \\
    $\alpha$-L-rhamnose & C1  &  0.290 &  0.091943 \\
    $\alpha$-L-rhamnose & C2  &  0.140 &  0.164784 \\
    $\alpha$-L-rhamnose & C3  &  0.140 &  0.245591 \\
    $\alpha$-L-rhamnose & C4  &  0.090 &  0.028042 \\
    $\alpha$-L-rhamnose & C5  &  0.110 &  0.305403 \\
    $\alpha$-L-rhamnose & C6  & -0.270 & -0.390280 \\
    $\alpha$-L-rhamnose & O2  & -0.650 & -0.682910 \\
    $\alpha$-L-rhamnose & O3  & -0.650 & -0.688366 \\
    $\alpha$-L-rhamnose & O4  & -0.360 & -0.388791 \\
    $\alpha$-L-rhamnose & O5  & -0.400 & -0.453004 \\
    $\alpha$-L-rhamnose & H1  &  0.090 &  0.172951 \\
    $\alpha$-L-rhamnose & H2  &  0.090 &  0.088685 \\
    $\alpha$-L-rhamnose & H3  &  0.090 &  0.042562 \\
    $\alpha$-L-rhamnose & H4  &  0.090 &  0.158716 \\
    $\alpha$-L-rhamnose & H5  &  0.090 &  0.086243 \\
    $\alpha$-L-rhamnose & H61 &  0.090 &  0.107484 \\
    $\alpha$-L-rhamnose & H62 &  0.090 &  0.107484 \\
    $\alpha$-L-rhamnose & H63 &  0.090 &  0.107484 \\
    $\alpha$-L-rhamnose & HO2 &  0.420 &  0.450668 \\
    $\alpha$-L-rhamnose & HO3 &  0.420 &  0.425346 \\
    Methyl chain starting & CM  &  0.090 &  0.116232 \\
    Methyl chain starting & HM1 &  0.090 &  0.038712 \\
    Methyl chain starting & HM2 &  0.090 &  0.038712 \\
    Methyl chain starting & HM3 &  0.090 &  0.038712 \\
    Methyl chain ending & CM  & -0.027 & -0.044241 \\
    Methyl chain ending & HM1 &  0.090 &  0.092539 \\
    Methyl chain ending & HM2 &  0.090 &  0.092539 \\
    Methyl chain ending & HM3 &  0.090 &  0.092539 \\
    \hline
    \caption{Atomic partial charges of the gellan repeating unit and the chain terminal groups.}
    \label{tbl:q}
    \end{longtable}

\subsection{S3. Gellan inter-chains hydrogen bonds}
\begin{figure}[htbp]
\centering
\includegraphics[width=0.5\textwidth]{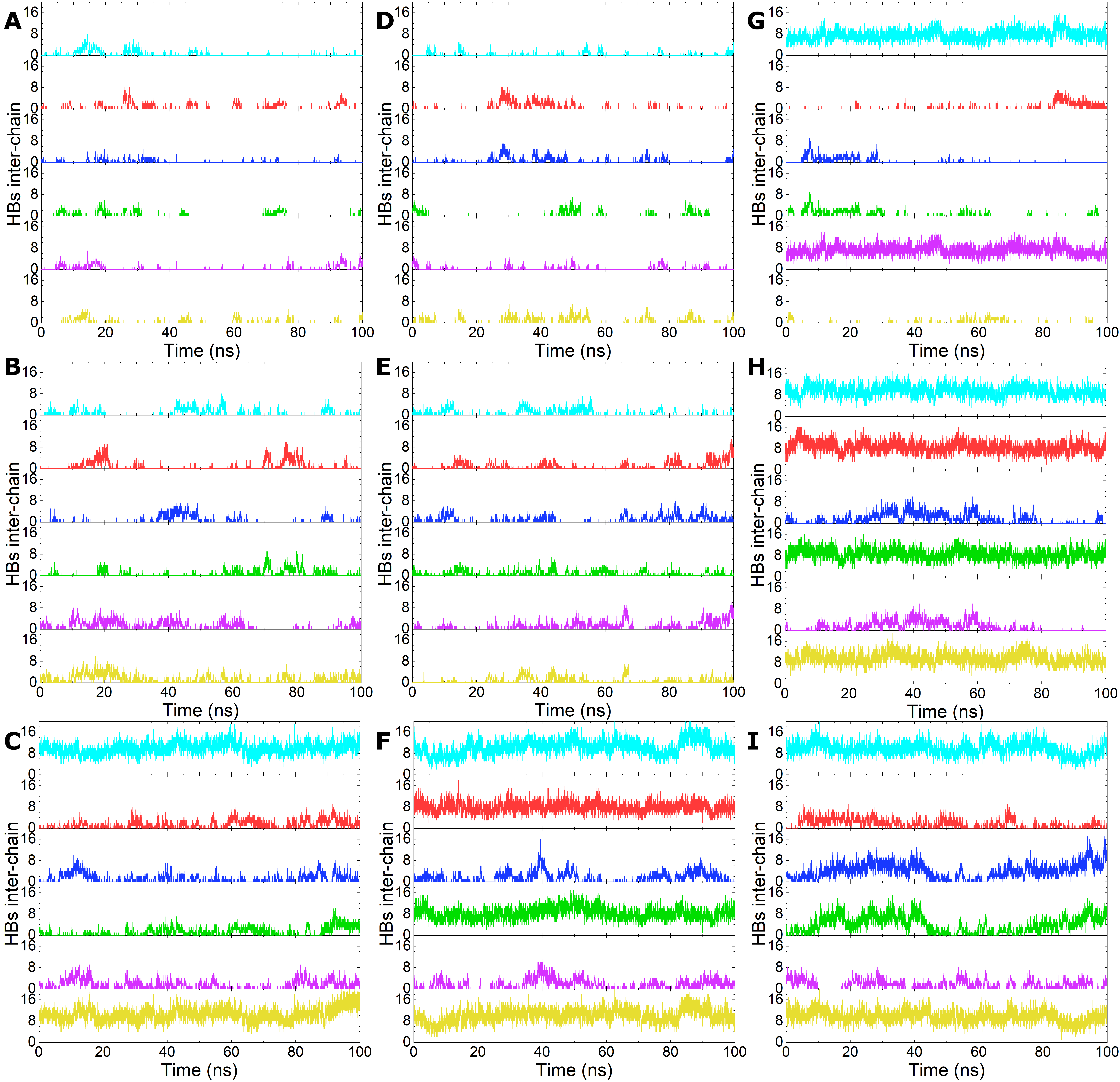}
  \caption{Time evolution of the total number of hydrogen bonds formed between a gellan chain with all the other chains from all-atom simulations at 298 K of pure gellan at concentration of 3 (A), 5 (B), and 10\%wt (C); gellan with sodium chloride 0.1 M at a polysaccharide concentration of 3 (D), 5 (E), and 10\%wt (F); and gellan with calcium acetate 0.05 M at a polysaccharide concentration of 3 (G), 5 (H), and 10\%wt (I). Data calculated for chain 1, 2, 3, 4, 5, and 6 are shown in cyan, red, blue, green, purple, and yellow, respectively.}
  \label{fgr:interHB}
\end{figure}

Hydrogen bonding interactions occurring between gellan chains were investigating by evaluating the time evolution of the number hydrogen bonds that each chain forms with all other chains. Figures~\ref{fgr:interHB}A-I show the results obtained for each of the six gellan chains in the different investigated systems. In all conditions, the number of hydrogen bonds between associated gellan chains appears stable with time.

\subsection{S4. Gellan chains conformation}
\begin{figure}[htbp]
\centering
\includegraphics[width=0.5\textwidth]{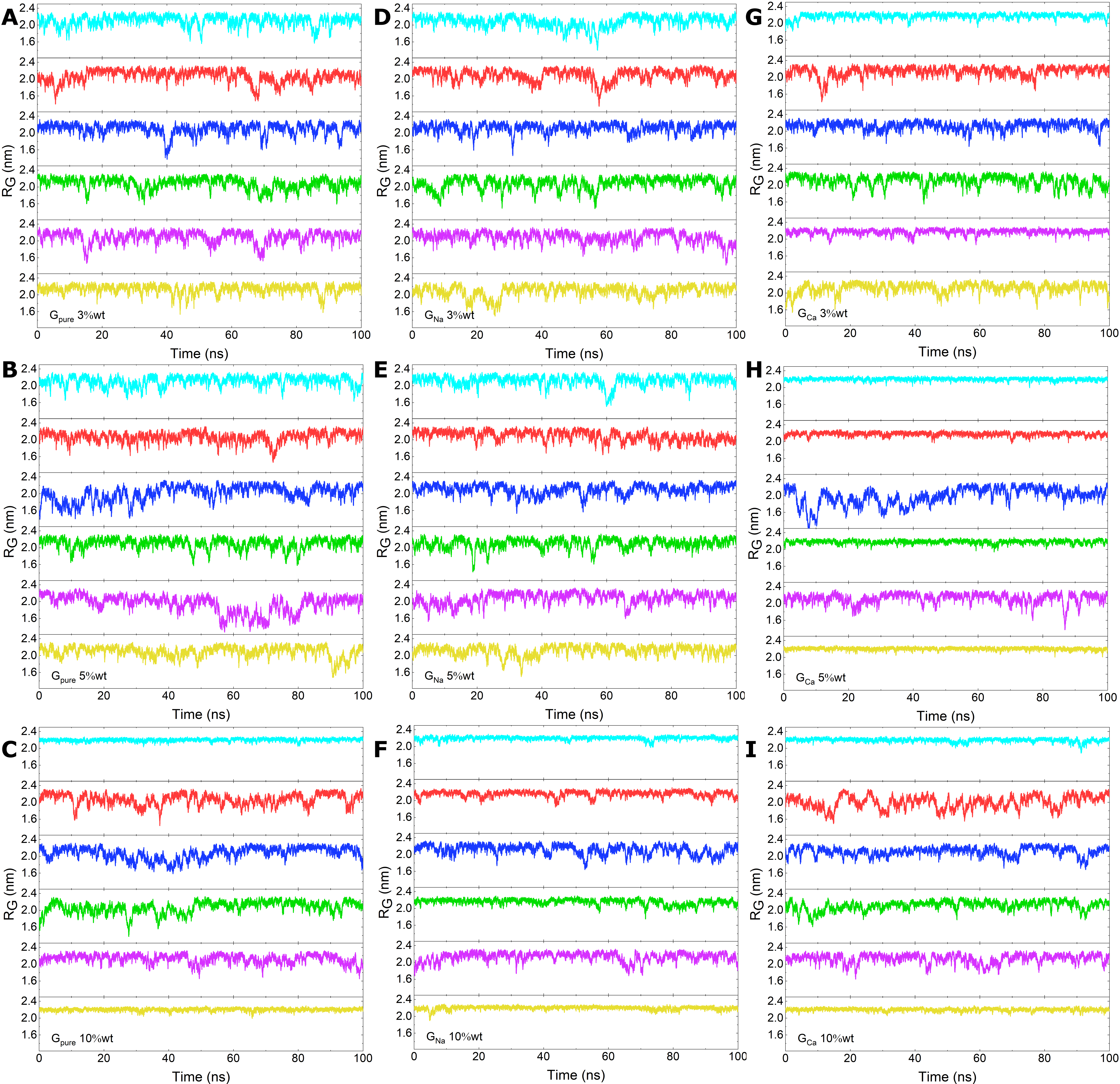}
  \caption{Time evolution of the radius of gyration of each gellan at 298 K calculated for pure gellan at concentration of 3 (A), 5 (B), and 10\%wt (C); gellan with sodium chloride 0.1 M at a polysaccharide concentration of 3 (D), 5 (E), and 10\%wt (F); and gellan with calcium acetate 0.05 M at a polysaccharide concentration of 3 (G), 5 (H), and 10\%wt (I). Data calculated for chain 1, 2, 3, 4, 5, and 6 are shown in cyan, red, blue, green, purple, and yellow, respectively.}
  \label{fgr:timerg}

\end{figure}

We have investigated the conformation of gellan chains by first monitoring the time evolution of the radius of gyration of each polysaccharide chain, as displayed in Figures~\ref{fgr:timerg}A-I. Aggregated gellan chains acquire rigid elongated conformations that exhibit a lower flexibility, as shown for the system $G_{Ca}$ 5\%wt (Figures~\ref{fgr:timerg}H) in which the fluctuations of the radius of gyration for aggregated chains (chains cyan and yellow) are considerably reduced as compared to the other chains (chains purple and blue).

\begin{figure}[htbp]
\centering
\includegraphics[width=0.5\textwidth]{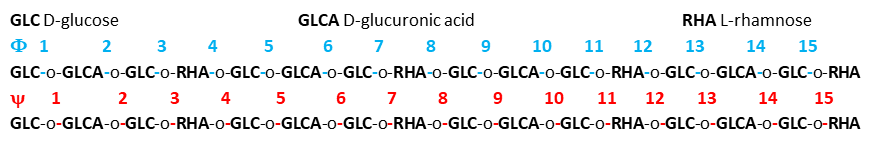}
  \caption{Schematic representation of the sequence of the glycosidic dihedral angles $\Phi$ (defined as H1-C1-O1-C4') and $\Psi$ (defined as C1-O1-C4'-H4') in a single gellan chain.}
  \label{fgr:dih}
\end{figure}

\begin{figure}[htbp]
\centering
\includegraphics[width=0.4\textwidth]{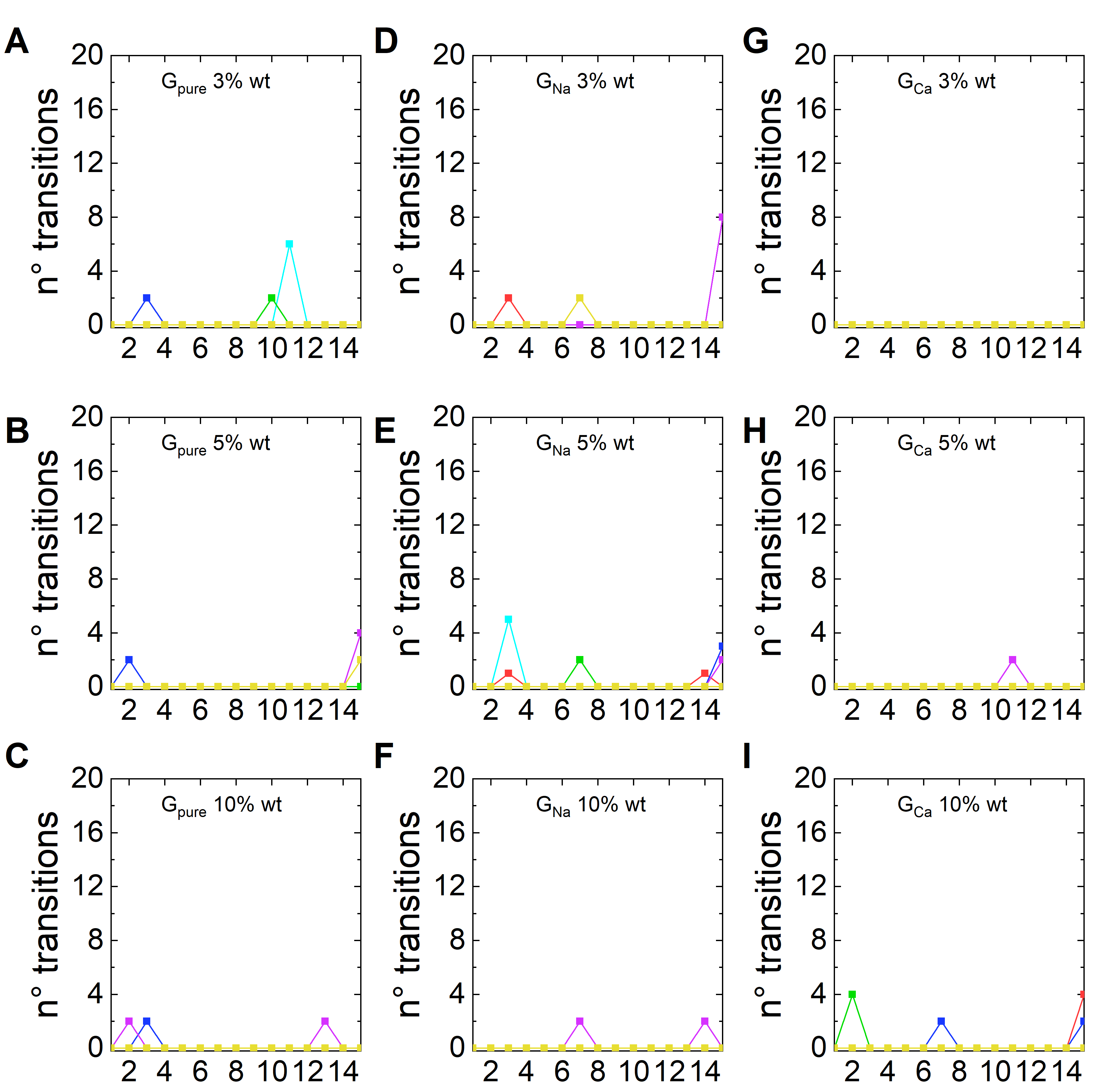}
  \caption{Number of dihedral transition of the glycosidic dihedral angle $\Phi$ defined as H1-C1-O1-C4' calculated over 100 ns for pure gellan at concentration of 3 (A), 5 (B), and 10\%wt (C); gellan with sodium chloride 0.1 M at a polysaccharide concentration of 3 (D), 5 (E), and 10\%wt (F); and gellan with calcium acetate 0.05 M at a polysaccharide concentration of 3 (G), 5 (H), and 10\%wt (I). Data calculated for chain 1, 2, 3, 4, 5, and 6 are shown in cyan, red, blue, green, purple, and yellow, respectively.}
  \label{fgr:fi}
\end{figure}

\begin{figure}[htbp]
\centering
\includegraphics[width=0.4\textwidth]{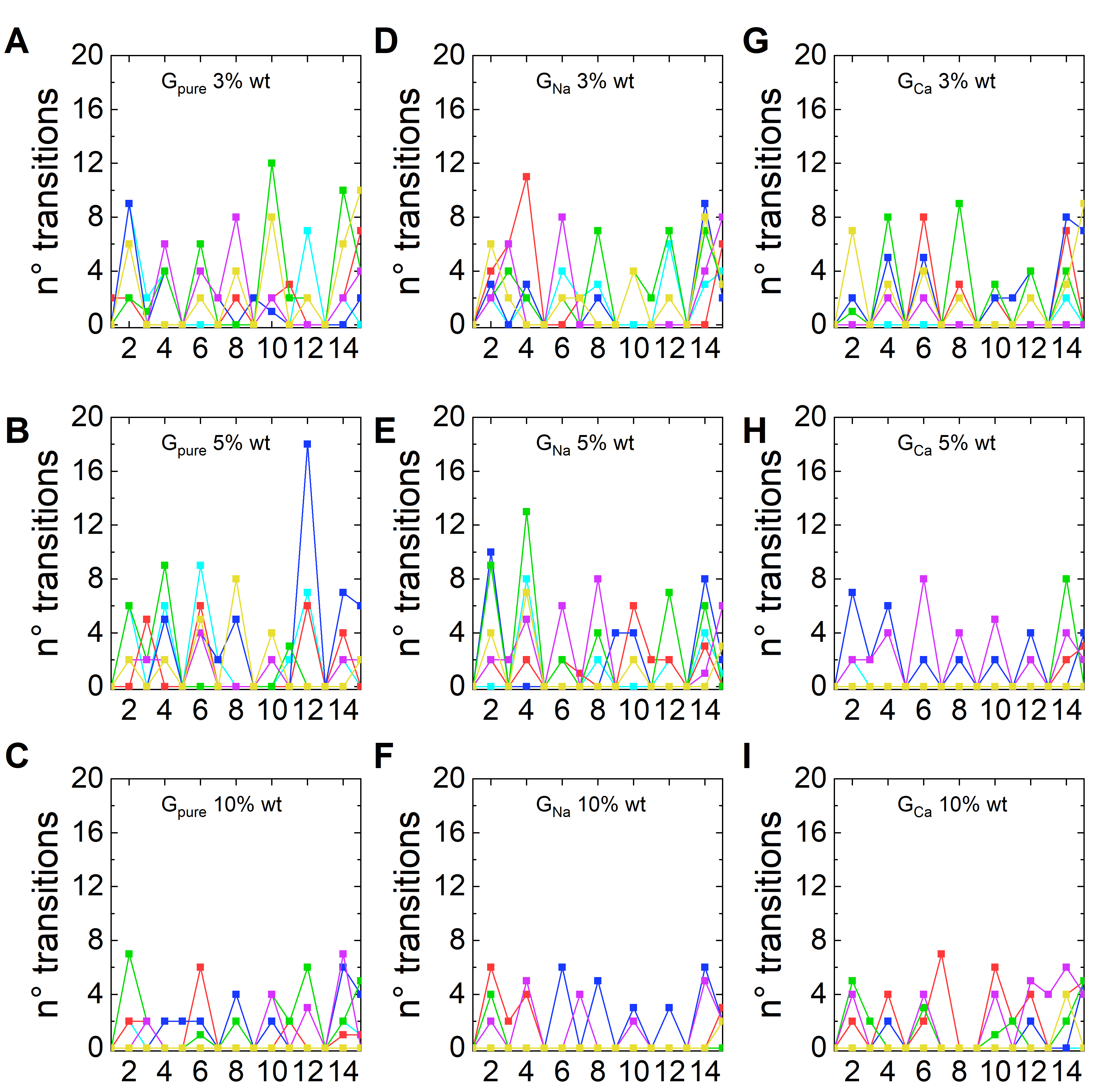}
  \caption{Number of dihedral transition of the glycosidic dihedral angle $\Psi$ defined as C1-O1-C4'-H4' calculated over 100 ns for pure gellan at concentration of 3 (A), 5 (B), and 10\%wt (C); gellan with sodium chloride 0.1 M at a polysaccharide concentration of 3 (D), 5 (E), and 10\%wt (F); and gellan with calcium acetate 0.05 M at a polysaccharide concentration of 3 (G), 5 (H), and 10\%wt (I). Data calculated for chain 1, 2, 3, 4, 5, and 6 are shown in cyan, red, blue, green, purple, and yellow, respectively.}
  \label{fgr:psi}
\end{figure}

Then, we have further investigated the flexibility of gellan chains by analyzing the behavior of the dihedral angles of the glycosidic linkages between the monosaccharides composing the chains. To this aim, we have evaluated the number of transitions of all the dihedral angles of each gellan chain $\Phi$, defined as H1-C1-O1-C4', and $\Psi$, defined as C1-O1-C4'-H4'. The occurrence of a dihedral transition was related to a dihedral angle change greater than 120$^{\circ}$. The sequence of glycosidic dihedral angles was defined as shown in the Scheme~\ref{fgr:dih}. Figures~\ref{fgr:fi}A-I and Figures~\ref{fgr:psi}A-I summarize the results obtained for the dihedral angle $\Phi$ and $\Psi$, respectively.
These findings show that the rotation of the dihedral angle $\Phi$ is quite restricted, independently on the polysaccharide concentration, presence of salt, or aggregation state of the chain. Moreover, for the dihedral angle $\Psi$, transitions are mostly detected for glycosidic linkages to glucose residues and are hindered for associated gellan chains, determining an increase of stiffness.

\end{document}